# Supervised Machine Learning Based Signal Demodulation in Chaotic Communications

Mykola Kozlenko

*Department of Information Technology*
*Vasyl Stefanyk Precarpathian National University*
Ivano-Frankivsk, Ukraine
https://orcid.org/0000-0002-2502-2447

*Abstract*—A chaotic modulation scheme is an efficient wideband communication method. It utilizes the deterministic chaos to generate pseudo-random carriers. Chaotic bifurcation parameter modulation is one of the well-known and widely-used techniques. This paper presents the machine learning based demodulation approach for the bifurcation parameter keying. It presents the structure of a convolutional neural network as well as performance metrics values for signals generated with the chaotic logistic map. The paper provides an assessment of the overall accuracy for binary signals. It reports the accuracy value of 0.88 for the bifurcation parameter deviation of 1.34% in the presence of additive white Gaussian noise at the normalized signal-to-noise ratio value of 20 dB for balanced dataset.

*Keywords*—deterministic chaos, chaotic signal, chaotic communications, bifurcation, bifurcation parameter keying, demodulation, bit error rate, machine learning, deep learning, convolutional neural network

## I. INTRODUCTION

Deterministic chaotic signals reflect the almost unpredictable dynamical behavior based on simple nonlinear rule [1]. Chaotic signals are successfully used as carriers in analog and digital communication systems [2], [3]. Chaotic modulation is an efficient wideband keying scheme [4]. It performs as a spread spectrum telecommunication system. Ergodic Chaotic Parameter Modulation (ECPM) is one of such keying schemes [5]. A modulation scheme determines how bits are mapped to the bifurcation parameter [6] of transmitted signals [7]. It requires sophisticated demodulation algorithms that increase the demodulator complexity. A logistic map [8] is a simple rule that can generate chaotic behavior. Logistic map based applications are widely used in various fields like digital communications, cybersecurity, etc. [9], [10]. This paper presents a bifurcation parameter demodulation approach for signals generated using the logistic map. It is based on supervised machine learning and considers communication in the presence of additive white Gaussian noise.

## II. RELATED WORK

Chaos has been applied in communication since 90s [11]. Paper [12] presents an overview of chaotic communications and basic chaotic modulation schemes like Chaos Shift Keying (CSK), Differential Chaos Shift Keying (DSK), Additive Chaos Modulation (ACM), and Multiplicative Chaos Modulation (MCM). It also describes synchronized chaotic systems and direct chaotic communication. Reference [13] proposes an information security inversion technology based on time series. It includes chaotic signal demodulation technology, chaotic channel modeling, and chaotic signal interference channel problem analysis. Paper [14] presents the particle filter to demodulate the Chaotic Parameter Modulation (CPM) signal. In [15] the state equations of unscented particle filter (UPF) for the demodulation of chaotic parameter have been derived. A novel modified algorithm based on UPF for the demodulation has been presented [15]. The papers [16] – [18] solve the problem of weak radio signals demodulation for Frequency Shift Keying (FSK) and Phase Shift Keying (PSK) signals with deep Machine Learning (ML) using convolutional neural networks (CNN) as well as fully connected networks. Paper [19] addresses the demodulating of underwater communication signal problem as a classification task and conceives a machine learning based demodulation scheme. Reference [20] presents the security weakness of communication method based on parameter modulation of a chaotic system and adaptive observer-based synchronization scheme. It has been shown that the security can be compromised without exact knowledge of the chaotic system properties [20]. Paper [21] presents the idea of using CNNs with deep learning structure to predict future symbols based on the received signal, to further reduce inter-symbol interference and to obtain a better Bit Error Rate (BER) performance in chaotic baseband wireless communication systems.

III. METHODOLOGY

Replacing the logistic equation

$$\frac{dx}{dt} = rx(1-x) \tag{1}$$

with the quadratic recurrence equation

$$x_{n+1} = rx_n(1-x_n), \tag{2}$$

where $r$ is a positive constant gives the logistic map [22]. Chaotic signals in this research are generated with the logistic map using Eq. 2. Fig. 1 presents the bifurcation diagram for $r$ in the range from 2.8 to 4.0.

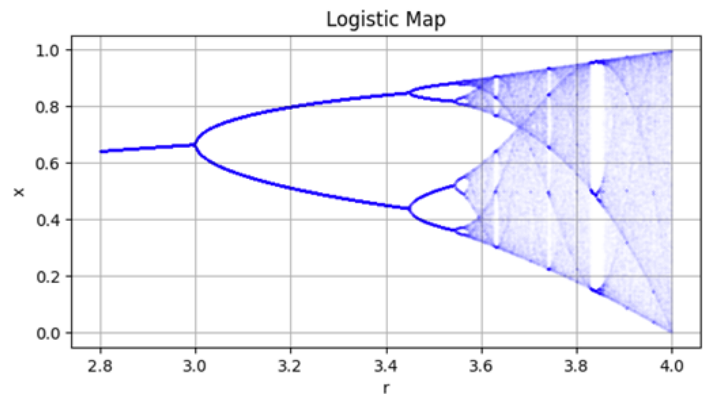

Fig. 1. Bifurcation diagram for the logistic map

Fig. 2 shows the time domain waveforms of the chaotic signals.

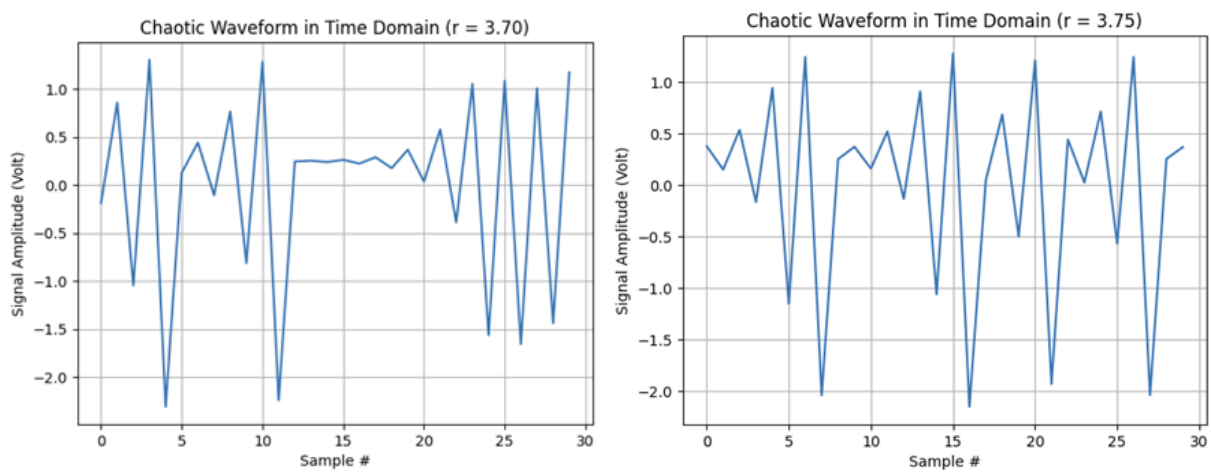

Fig. 2. Time domain waveforms of chaotic signals for parameter $r = 3.7$ and $r = 3.75$

2022 International Conference on Innovative Solutions in Software Engineering
Ivano-Frankivsk, Ukraine, November 29-30, 2022

In this research, the modulation scheme represents digital data as a variation in the bifurcation parameter of a chaotic baseband carrier wave. The deviation of the bifurcation parameter value is 0.05; from 3.7 for "space" to 3.75 for "mark" (percent deviation equals 1.34%). The baseband bandwidth is 5512.5 Hz. The sampling frequency is 11025 Hz. The bit duration time is 0.3715 s, 4096 samples (exactly as it is in JT65 protocol). The bit rate is 2.67 bit/s. The signal-to-noise ratio (SNR) is -13 dB, the normalized SNR ($E_b/N_0$) is +20 dB, and the spreading factor is +33 dB).

Fig. 3 shows the structure of the convolutional neural network that is used for the demodulation.

```
Model: "sequential"
_________________________________________________________________
Layer (type)                 Output Shape              Param #
=================================================================
reshape (Reshape)            (None, 4096, 1)           0
_________________________________________________________________
batch_normalization (BatchNo (None, 4096, 1)           4
_________________________________________________________________
conv1d (Conv1D)              (None, 4096, 128)         2176
_________________________________________________________________
batch_normalization_1 (Batch (None, 4096, 128)         512
_________________________________________________________________
flatten (Flatten)            (None, 524288)            0
_________________________________________________________________
dense (Dense)                (None, 64)                33554496
_________________________________________________________________
batch_normalization_2 (Batch (None, 64)                256
_________________________________________________________________
dense_1 (Dense)              (None, 2)                 130
=================================================================
Total params: 33,557,574
Trainable params: 33,557,188
Non-trainable params: 386
```

Fig. 3. Structure of the convolutional neural network

The size of the training dataset is 12800 records, the validation set size is 3200, and the test set size is 4000 records. Loss function: categorical cross-entropy, optimizer: adam. Two output neurons are used in the model and labels are one-hot encoded. Thus, the model can be easily converted to a multiclass classifier for m-ary signals.

IV. RESULTS

The classification report for the test data is shown in Table. 1.

TABLE 1. CLASSIFICATION REPORT

| Class | Classification Metrics | | | |
|---|---|---|---|---|
| | Precision | Recall | F1-Score | Support |
| 0 – "space" | 0.94 | 0.82 | 0.87 | 2003 |
| 1 – "mark" | 0.84 | 0.95 | 0.89 | 1997 |
| Accuracy | | | 0.88 | 4000 |
| Macro avg | 0.89 | 0.88 | 0.88 | 4000 |
| Weighted avg | 0.89 | 0.88 | 0.88 | 4000 |

## V. Discussion

The purpose of this study was to gain a better understanding of the ability of machine learning algorithms to demodulate and decode the baseband chaotic signals. The result of the present study supports the hypothesis that chaotic signal generated with the logistic map can be demodulated with ML techniques. The results of this research provide supporting evidence that it is possible even in the presence of noise. This is the main take away from this paper. This pattern of results is consistent with the previous works [16], [18] those deal with ML-based demodulation for frequency shift keying modulation scheme. These results represent the direct demonstration of chaotic signal demodulation within the scope of the ML approach. There are at least three potential limitations concerning the results of this study. A first limitation is that we used only signals generated with the logistic map. A second potential limitation is that we used supervised machine learning approach. This means that the algorithm should be trained with previously known signal patterns. Unsupervised approach looks like more promising technique. The third limitation is that we considered only the case of AWGN. Despite these limitations, the results suggest practical implication that chaotic signals can be demodulated without exact knowledge of the bifurcation parameter values.

## VI. Future research

In terms of future research, it would be useful to extend the current findings by examining other types of chaotic maps. Also, it would be useful to research the ability of unsupervised ML approach in the context of this task. Other types of interferences should be studied as well.

## VII. Conclusion

The main conclusion that can be drawn is that chaotic baseband signals generated with the logistic map in the bifurcation parameter modulation scheme can be demodulated using supervised machine learning approach. In summary, this paper presents the values of quality metrics. The overall accuracy is 0.88 at normalized SNR value of +20 dB.

## VIII. Acknowledgment

The author gratefully acknowledges the contributions of scientists of the Department of Information Technology of the Vasyl Stefanyk Precarpathian National University for scientific guidance given in discussions and technical assistance helped in the actual research.

## IX. Ethics declarations

The author has nothing to disclose.